\documentclass[%
 reprint,
 superscriptaddress,
 amsmath,amssymb,
 aps,
 prb,
]{revtex4-1}

\usepackage[utf8]{inputenc}
\usepackage{booktabs}
\usepackage[english]{babel}

\AtBeginDocument{
\heavyrulewidth=.08em
\lightrulewidth=.05em
\cmidrulewidth=.03em
\belowrulesep=.65ex
\belowbottomsep=0pt
\aboverulesep=.4ex
\abovetopsep=0pt
\cmidrulesep=\doublerulesep
\cmidrulekern=.5em
\defaultaddspace=.5em
}

\usepackage{graphicx}
\usepackage{dcolumn}
\usepackage{bm}
\usepackage[nomargin,inline,marginclue,draft]{fixme}
\usepackage{hyperref}
\usepackage{algorithm}
\usepackage{algpseudocode}
\usepackage{siunitx}
\usepackage[caption=false]{subfig}

\makeatletter
 \begingroup
 \catcode`\_=\active
 \protected\gdef_{\@ifnextchar|\subtextup\sb}
 \endgroup
 \def\subtextup|#1|{\sb{\textup{#1}}}
 \AtBeginDocument{\catcode`\_=12 \mathcode`\_=32768 }
\makeatother

\newcommand{\figref}[1]{\figurename~\ref{#1}}

\begin{document}

\preprint{APS/123-QED}

\title{Materials property prediction using symmetry-labeled graphs as atomic-position independent descriptors}

	\author{Peter Bjørn Jørgensen}%
	\affiliation{
	 Department of Energy Conversion and Storage, Technical University of Denmark
	}%
	\author{Estefanía Garijo del Río}
	\affiliation{
	 Department of Physics, Technical University of Denmark
	}%
	\author{Mikkel N. Schmidt}
	\affiliation{
	 Department of Applied Mathematics and Computer Science, Technical University of Denmark
	}%
	\author{Karsten Wedel Jacobsen}
	\affiliation{
	 Department of Physics, Technical University of Denmark
	}%
%
%
%
%
%
\begin{abstract}
Computational materials screening studies require fast calculation of the properties of thousands of materials. The calculations are often performed with Density Functional Theory (DFT), but the necessary computer time sets limitations for the investigated material space. Therefore, the development of machine learning models for prediction of DFT calculated properties are currently of interest. A particular challenge for \emph{new} materials is that the atomic positions are generally not known. We present a machine learning model for the prediction of DFT-calculated formation energies based on Voronoi quotient graphs and local symmetry classification without the need for detailed information about atomic positions. The model is implemented as a message passing neural network and tested on the Open Quantum Materials Database (OQMD) and the Materials Project database. The test mean absolute error is 22 meV on the OQMD database and 43 meV on Materials Project Database. The possibilities for prediction in a realistic computational screening setting is investigated on a dataset of 5976 ABSe$_3$ selenides with very limited overlap with the OQMD training set. Pretraining on OQMD and subsequent training on 100 selenides result in a mean absolute error below 0.1 eV for the formation energy of the selenides.
\end{abstract}
%
\maketitle
\section{Introduction}

Over the last decades, high-throughput computational screening studies
have been employed to identify new materials within different areas such as
(photo-)electrochemistry \cite{Castelli:2012,Wu:2012ce,kuhar:2017}, batteries \cite{Urban:2016gn,Aykol2016}, catalysis \cite{Andersson:2006iv, Norskov:2009je}, and more \cite{Mounet:2018ks, Curtarolo:2013fa, Alberi:2018dm}. Such
studies are typically based on Density Functional Theory \cite{Hohenberg:1964fz, Kohn:1965js} and because
of computational requirements they are usually limited to some
thousands or tens of thousands of materials. In order to investigate
larger parts of the huge space of possible materials, new methods are
needed to perform faster calculations or to guide the search in the
material space in a more informed way.

One way to circumvent the computationally demanding DFT calculations
is to use machine learning (ML) techniques to predict materials
properties, and this approach has been explored intensively the last
years. Several descriptors or fingerprints to characterize the atomic
structure of a material have been suggested including the partial radial distribution function \cite{Schutt:2014fb} and the Coulomb matrix \cite{Faber:2015fx}. More involved fingerprints combining many atomic properties
and crystal structure attributes based on Voronoi graphs have also been developed \cite{Ward2017-ff, Oses:2017cd}, along with graph representations, which are directly mapped onto convolutional neural networks \cite{Schutt2017-hk, Schutt2018-ik, Xie:2018ia}.

The use of ML to speed up DFT calculations may have several goals in a computational screening setting. If the atomic structure (i.e. the positions of all the atoms) of a material is known, ML may in principle provide the same information about the material as a DFT calculation would: structural stability, phonon dispersion relations, elastic constants etc. It might even in principle provide data of a better quality than standard (semi-)local DFT calculations, comparable to more advanced DFT calculations with hybrid functionals or even higher-level methods as recently demonstrated for molecules \cite{Zaspel:2019jn}.

However, the atomic positions of \emph{new} materials will generally not be known. If the atomic positions are known from experiment, the material is not really new (even though many of its properties might be unknown) and if the positions are obtained from a DFT calculations there is no need to use a ML prediction of already calculated properties.

Our focus here will be the prediction of properties of \emph{new} materials where the detailed atomic positions are unknown, and since the most crucial property of a new material is its stability we shall concentrate on prediction of formation energies.

The obvious question of course then is, how we can describe or classify a crystalline material without  knowing  the explicit positions of the atoms. The most fundamental property of a material is its chemical composition, i.e. for a ternary material $A_x B_y C_z$, the identity of the elements $A$, $B$, and $C$ and their relative appearance $x:y:z$. It turns out that based on this information alone a number of predictions about material stability can be made. Meredig et al. \cite{Meredig2014-or} demonstrated that it is possible to predict thermodynamic stability of new compounds with reasonable accuracy based on composition alone, and a number of new compound compositions were predicted and their structures subsequently determined. However, this approach of course has its limitations as it cannot distinguish between materials with the same composition but different crystal structures.

A rigorous classification of a crystalline material comes from its symmetry. Any periodic material belongs to one of the 230 space groups, and this puts restrictions on the possible atomic positions. In the simplest cases of, say, a unary material with one atom in the unit cell with space group Fm-3m (an fcc crystal), all atomic positions are determined up to a scaling of the volume. Similarly, the fractional positions (i.e. relative to the unit cell) of the atoms in materials with several elements can be determined entirely by symmetry as for example shown for BaSnO$_3$ in the cubic perovskite structure in Figure~\ref{fig:basno3_atoms}. More generally, scaled atomic positions may be fully or partially determined depending on their symmetry, and the symmetry properties can be expressed using the so-called Wyckoff sites. This classification was recently used by Jain and Bligaard \cite{Jain2018-df} to build a machine learning model based on only composition and the Wyckoff positions, i.e. without any detailed information about the atomic positions. They were able to achieve a  mean absolute errors of about 0.07 eV/at on the prediction of the formation energy on a test dataset of more than 85000 materials.

Here, we shall develop a machine learning model, which does not require knowledge of the detailed atomic positions. However, unlike the model proposed by Jain and Bligaard, it will be based on local information about interatomic bonds and the symmetry of their environments. The bonds will be identified using Voronoi graphs and the symmetry will be classified using the Voronoi facets. The resulting model has a mean absolute error on the heats of formation for the OQMD database of only 22 meV  and for the ICSD part of OQMD it is 40 meV.

In section \ref{sec:representation} we describe the proposed graph representation based on quotient graphs and the classification of Voronoi facet point symmetry and in section \ref{sec:graph_prototypes} we investigate the relation between quotient graphs and prototypes based on data from OQMD. This is followed by an introduction of the machine learning model and the datasets in section \ref{sec:neural_message_passing} and \ref{sec:datasets} respectively. The numerical results are presented in section \ref{sec:results} and followed by the conclusions in section \ref{sec:conclusions}.

\section{Graph Representation}\label{sec:representation}
As representation for the machine learning algorithm we use the quotient graph as introduced by \cite{Chung1984-pb} and also used in \cite{Xie:2018ia}.
The quotient graph is a finite graph representation of the infinite periodic network of atoms.
Every atom in the unit cell corresponds to a vertex of the quotient graph. We denote the graph $G$ and the set of $N$ vertices $\left\{ v_i \right\}_{i=1}^{N}$.
When two atoms are connected in the network we draw an edge between the atoms in the quotient graph.
In this work we use the Voronoi diagram to decide when two atoms are connected, specifically a pair of atoms are connected if they share a facet in the Voronoi diagram.
Due to periodic boundary conditions a pair of atoms may share several facets and in this case there will be several edges between the atoms.
When interatomic distances are available the edges are labeled with the distance between the atoms.

\begin{figure}[tbp]
	\centering\includegraphics[width=0.7\columnwidth]{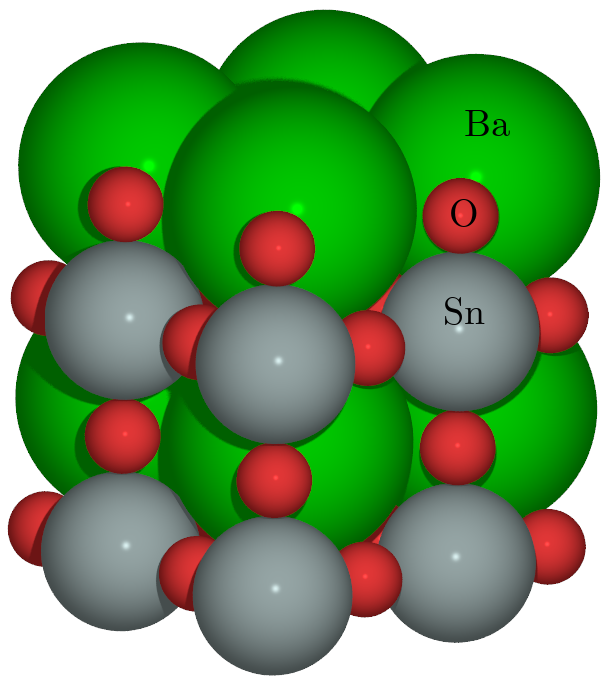}
	\caption{Structure of BaSnO3. The unit cell contains one Ba atom (green), one Sn atom (grey) and three O atoms (red).}
	\label{fig:basno3_atoms}
\end{figure}

\begin{figure}[tbp]
	\centering\includegraphics[width=0.7\columnwidth]{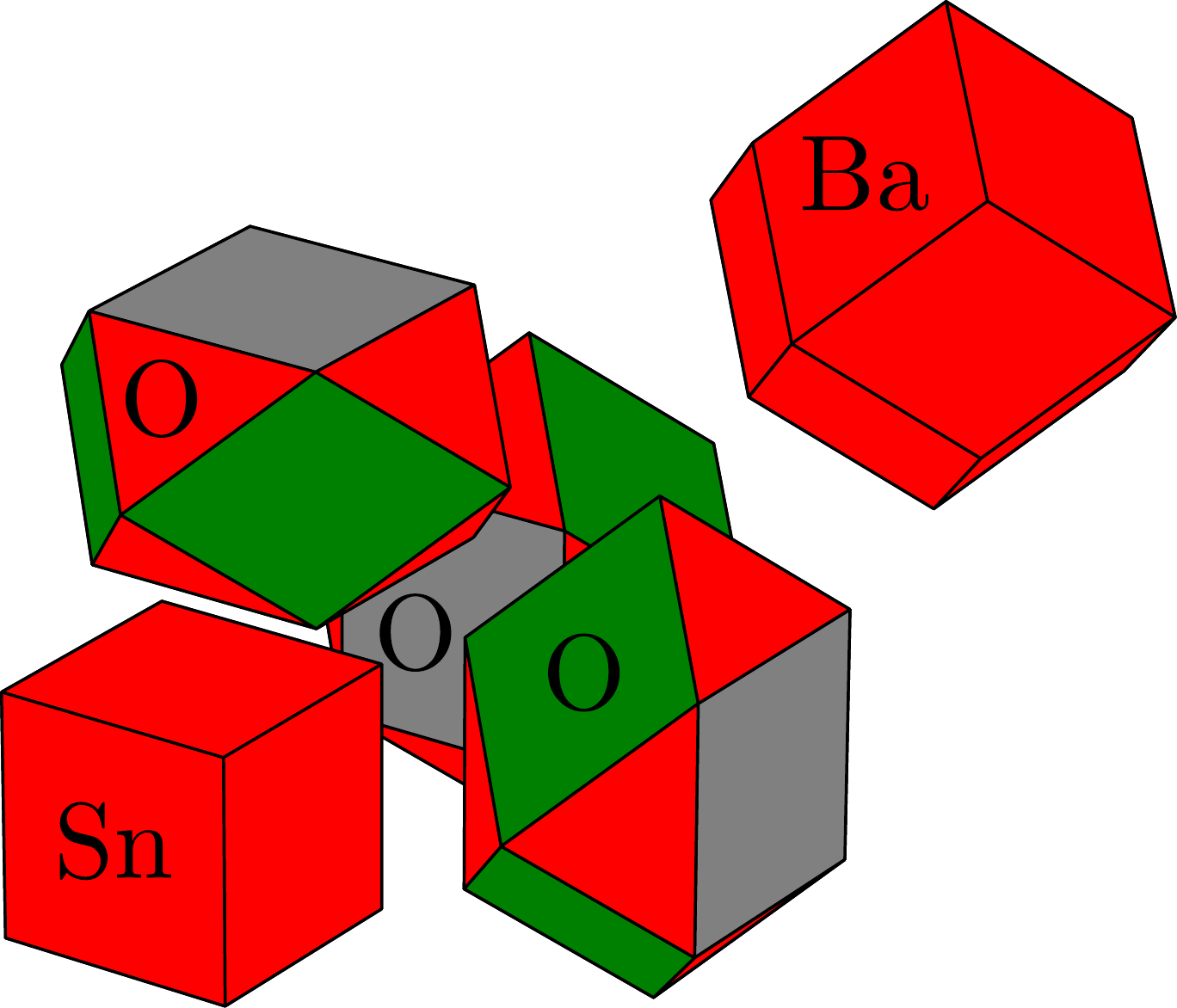}
	\caption{Voronoi cells of BaSnO3. The cells have been displaced for the visualization. The color of the facets corresponds to the atomic species of the neighbouring atom (green for Ba, grey for Sn and red for O neighbours).}
	\label{fig:basno3_cells}
\end{figure}

\begin{figure}[tbp]
	\centering\includegraphics[width=1.0\columnwidth]{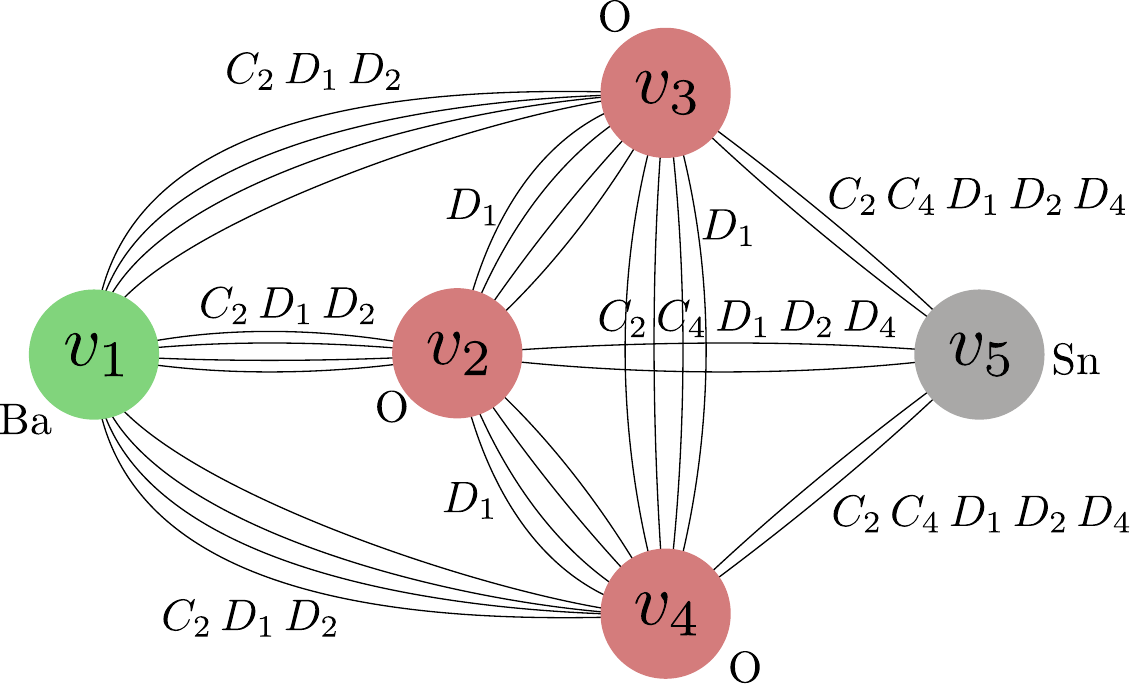}
	\caption{Quotient graph for BaSnO3. The edge labels show the point groups of the corresponding facets of the Voronoi diagram. For this particular case, the repeated edges between vertices all have the same point groups, but in general they could be different.}
	\label{fig:basno3_graph}
\end{figure}

As an example we look at BaSnO3 in the perovskite structure as shown in \figref{fig:basno3_atoms}.
This material has five atoms in the unit cell.
After performing Voronoi tessellation we get a Voronoi cell for each atom in the unit cell as shown in \figref{fig:basno3_cells}.
The Voronoi diagram defines the edges in the quotient graph which is illustrated in \figref{fig:basno3_graph}. 

An inherent problem with Voronoi graph construction method is that small perturbations of the atom positions may lead to different graphs. Classification of different types of instabilities has even been used by \citet{Lazar2015-wb} to characterize local structure.
As shown by \citet{Reem2011-cs} small changes in the Voronoi sites lead to only small changes in the Voronoi cell volume.
However, small perturbations can still lead to appearances of quite small facets. This is for example often the case for structures with high symmetry, where small displacements of the atoms introduce new facets. To increase the stability we remove these small facets and the corresponding connections in the graph by introducing a cutoff in the solid angle of the facet $\Omega_{cut}$. We use $\Omega_{cut} = 0.2$, but as we shall see later the results are surprisingly stable with regards to increasing this value. A more advanced method for improving the stability of the Voronoi graph has been proposed by \citet{Malins2013-ti}.

The graph is annotated with the symmetry group of each of the Voronoi facets.
In the following section we describe this symmetry classification in more detail.


\subsection{Symmetry Group Classification}\label{sec:sym_classification}
\begin{figure}[tbp]
	\centering\includegraphics[width=0.5\columnwidth]{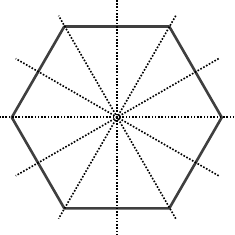}
	\caption{Mirror axes of a hexagon.}
	\label{fig:polygon_mirror_axes}
\end{figure}

\begin{figure*}[tbp]
	\centering\includegraphics[width=0.8\textwidth]{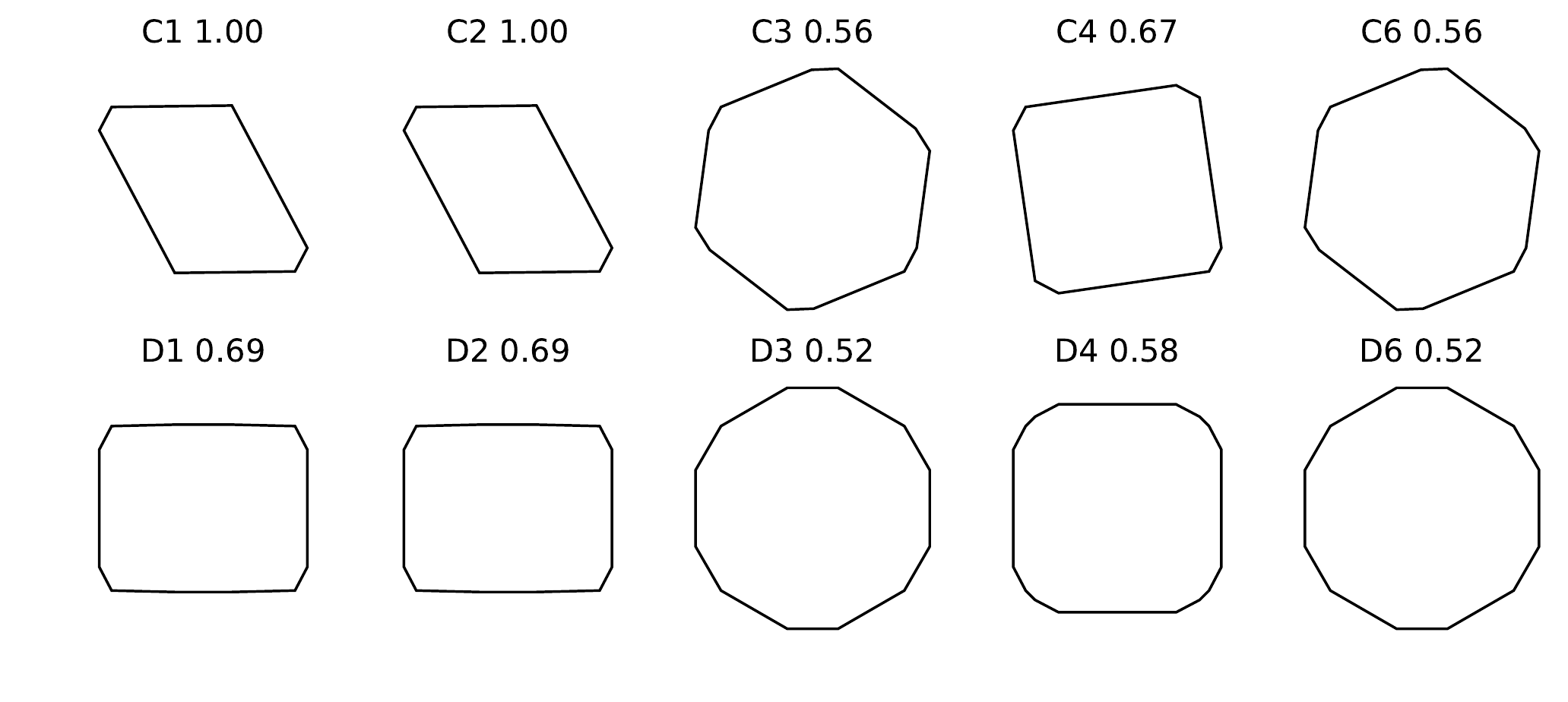}
	\caption{Each shape is the convex hull of the shape in the top left corner after the symmetry operations of each of the point groups have been applied. The label above each shape denotes the point group and the symmetry measure for that group.}
	\label{fig:symmetry_shapes}
\end{figure*}

To characterize the symmetry of an atomic environment we classify the  symmetry of each Voronoi facet into the 9 non-trivial two-dimensional point groups $(C_2,C_3,C_4,C_6, D_1, D_2, D_3, D_4, D_6)$.
The classification method is inspired by the symmetry measure introduced by \citet{heijmans}. Given the vertices of the two-dimensional Voronoi facet we go through the following procedure
\begin{enumerate}
    \item Compute centroid and center the shape.
    \item Search for mirror axis and align it with the x-axis if it exists. \label{step:mirror_search}
    \item For each point symmetry group apply all elements of the group and calculate the area of the convex hull of the new points generated by this procedure.
\end{enumerate}
The symmetry measure is then the ratio between the area of the original shape and the area defined by the convex hull of the new vertices.
When the symmetry measure for a given group is close to unity we label the facet as having this symmetry.
See \figref{fig:symmetry_shapes} for an example shape and its symmetry measure for each group.
The search for mirror axis in step \ref{step:mirror_search} is done by computing the moment of inertia and test the two principal axes for mirror symmetry. When the moments of inertia are the same, for example when the shape is a regular polygon, the principal axes are arbitrary and we fall back to testing for mirror symmetry at all axes going through the centroid and either a vertex or a midpoint of a line segment. For a regular hexagon these axes are illustrated in \figref{fig:polygon_mirror_axes}.

\section{Graph representation and prototypes}\label{sec:graph_prototypes}

In many applications prototypes are used as a descriptor for the overall structure of a material and as part of a computational screening procedure some of the atoms of the prototypes may be swapped with other elements.
We want to assess whether there is a correspondence between the prototypes and Voronoi graphs, i.e. do two materials with the same prototype have the same Voronoi graph and do two materials with the same Voronoi graph have the same prototype?
The question cannot be ultimately answered because prototype naming is not completely well-defined: in some cases several different prototypes are used to describe the same material, and many materials may not have prototypes attached to them. But we can show to which extent Voronoi graphs is aligned with the use of prototypes.

For this analysis we use the OQMD database with the prototypes assigned in the database. We note that this assignment is not generally unique. For example an elemental compound in the fcc structure may be labeled with either "Cu" or "A1\_Cu" in the database. In other cases two clearly different structures are classified with the same prototype.

We investigate all unary, binary and ternary compounds in the database and for each of these sets we study the link between graphs $G$ and prototypes $P$, i.e. if we know that a given structure has a specific prototype do we then also know which graph it has and vice versa.
One way of measuring this is through the mutual information $I(G;P)$ of $G$ and $P$.
The mutual information is symmetric and can be computed as
\begin{align}
	I(G;P) &= H(G) - H(G|P) \\
		& = H(P) - H(P|G),
	\label{eq:mutual_info}
\end{align}
where $H$ denotes the entropy. The mutual information is thus the average decrease in entropy we get from knowing the other variable. We also compute the normalized mutual information known as the uncertainty coefficient $U(X|Y)=I(X;Y)/H(X)$ which can be seen as given $Y$ what fraction of bits of $X$ can we predict.
To compute these quantities we need the distribution over graphs and we obtain these distributions approximately by comparing graph fingerprints.\footnote{The graph fingerprints are computed using the neural message passing model with random weight initialization. We use two instances of neural network weight initialization and six different atomic embedding instances, thus having 12 models in total. The fingerprint is then a vector where each entry is the scalar output of one of these models.}
The quantities for OQMD are shown for the unlabeled graph in \tablename~\ref{tab:graph-prot} and for the graph labeled with rotation symmetries in \tablename~\ref{tab:graph-prot-wsym}.

The uncertainty coefficient is close to 90\% in most cases except for the Unary compounds $U(P|G)$. In this case structures with different prototypes map to the same graph and we may be discarding important structural information.
Including symmetry information increases the number of unique graphs significantly, which implies that the uncertainty coefficient $U(G|P)$
decreases while $U(P|G)$ increases.

\begin{table*}
	\subfloat[][for graphs without symmetry labels]{
	\centering
	\begin{tabular}{lrrrrrrrr}
		\toprule
		& $N$ & $|G|$ & $|P|$ & $H(G)$ & $H(P)$ & $I(G;P)$ & $U(G|P)$ & $U(P|G)$ \\
		\midrule
		Unary 				& 1487	& 85	&   67&  4.4 & 4.7  & 3.7  	&  0.84	& 0.80	\\
		Unary \sc{icsd}		& 196	& 46	&	49&	 4.2 & 4.2	& 3.8  	&  0.90	& 0.90 \\
		Binary 				& 53528	& 1318	&  871&  4.3 & 4.5  & 3.8  	&  0.90 & 0.86	\\
		Binary \sc{icsd}	& 5862	& 1219	&  850&  8.2 & 8.0  & 7.6  	&  0.92 & 0.95	\\
		Ternary 			& 339960& 4006	& 1754&  2.0 & 1.9  & 1.8  	&  0.91 & 0.98 \\
		Ternary \sc{icsd} 	& 11500 & 3487	& 1740&  10.0& 9.1 	& 8.8	&  0.88 & 0.97	\\
		\bottomrule
	\end{tabular}
	\label{tab:graph-prot}}

	\subfloat[][for graphs with symmetry labels]{
	\centering
	\begin{tabular}{lrrrrrrrr}
		\toprule
		& $N$ & $|G|$ & $|P|$ & $H(G)$ & $H(P)$ & $I(G;P)$ & $U(G|P)$ & $U(P|G)$ \\
		\midrule
		Unary 				& 1487	& 222	&   67&  6.1 & 4.7  & 4.3  &  0.70   & 0.91	\\
		Unary \sc{icsd}		& 196	& 68	&   49&  4.8 & 4.2  & 4.0  &  0.84   & 0.96	\\
		Binary 				& 53528	& 2040	&  871&  4.7 & 4.5  & 4.0  &  0.84   & 0.90	\\
		Binary \sc{icsd}	& 5862	& 1742	&  850&  8.8 & 8.0  & 7.8  &  0.89   & 0.97	\\
		Ternary 			& 339960& 5703	& 1754&  2.1 & 1.9  & 1.8  &  0.88   & 0.99 \\
		Ternary \sc{icsd} 	& 11500 & 4504	& 1740& 10.5 & 9.1  & 9.0  &  0.85   & 0.99	\\
		\bottomrule
	\end{tabular}
	\label{tab:graph-prot-wsym}}
	\caption{Correspondence between Voronoi graphs and prototypes in OQMD  with and without symmetry labels. $N$ denotes the number of entries, $|G|$ the number of unique Voronoi graphs and $|P|$ the number of different prototypes. $H(G)$ and $H(P)$ are the entropy of the distribution of graphs and prototypes respectively, while $I(G;P)$ is the mutual information between the two distributions and $U(G|P)$, $U(P|G)$ are the normalized mutual information.}
\end{table*}

\section{Neural Message Passing Model}\label{sec:neural_message_passing}
In this section we introduce the machine learning model which takes the labeled graph as input and outputs an energy prediction as a scalar.
We describe the model as message passing on a graph following the notational framework introduced by \citet{Gilmer2017-wp}.
We follow the message passing notation but the model we are going to introduce can be seen as an extension of the SchNet model \cite{Schutt2018-ik}, which can also be cast into this framework as we have shown in prior work \cite{Jorgensen2018-op}.

Denote the graph $G$ with vertex features $x_v$ and edge features $\varepsilon_{vw}$ for an edge from vertex $v$ to vertex $w$.
Each vertex has a hidden state $h_v^t$ at ``time'' (or layer) $t$ and we denote the edge hidden state $e_{vw}^t$.
The hidden states of vertices and edges are updated in a number of interaction steps $T$.
In each step the hidden state of vertices are updated in parallel by receiving and aggregating messages from neighbouring vertices.
The messages are computed by the message function $M_t(\cdot)$ and the vertex state is updated by a state transition function $S_t(\cdot)$, i.e.
\begin{align}
	m_{v}^{t+1} &= \sum_{w \in N(v)} M_t\left( h_v^t, h_w^t, e_{vw}^t \right),\\
	h_{v}^{t+1} &= S_{t}\left( h_v^t, m_v^{t+1} \right),
	\label{eq:message_function}
\end{align}
where $N(v)$ denotes the neighborhood of $v$, i.e. the vertices in the graph that has an edge to $v$.
The edge hidden states are also updated by an edge update function $E_t(\cdot)$ that depends on the previous edge state as well as the vertices that the edge connects:
\begin{equation}
	e_{vw}^{t+1} = E_t(h_v^t, h_w^{t}, e_{vw}^t).
	\label{eq:edge_update}
\end{equation}
After $T$ interaction steps the vertex hidden state represents the atom type and its chemical environment.
We then apply a readout function $R(\cdot)$ which maps the set of vertex states to a single entity
\begin{equation}
	\hat{y} = R\left( \left\{ h_v^T \in G \right\} \right)
	\label{eq:readout}
\end{equation}
The readout function operates on the set of vertices and must be invariant to the ordering of the set.
This is often achieved simply by summing over the vertices.
In some architectures the final edge states are also included as an argument to the readout function.
The message function $M_t(\cdot)$, state transition function $S_t(\cdot)$, edge update function $E_t(\cdot)$ and readout function $R(\cdot)$ are implemented as neural networks with trainable weight matrices.
To fully define the model we just need to define these functions and a number of models can be cast into this framework.
We use different weight matrices for each time step $t$, however in some architectures the weights are shared between layers to reduce the number of parameters.

In this work we use the model proposed in our prior work \cite{Jorgensen2018-op}. The model is an extension of the SchNet model \cite{Schutt2018-ik}, with the addition of an edge update network.
The message function  is only a function of the sending vertex and can be written as
\begin{align}
	M_{t}(h_w^t, e_{vw}^t) &=  (W_{1}^{t} h_w^{t}) \odot g(W_{3}^{t}g(W_{2}^{t}e_{vw}^t)),
\end{align}
where $\odot$ is element-wise multiplication, the $W$'s are weight matrices and $g(x)$ is the activation function, more specifically the shifted soft-plus function $g(x)=\ln( \operatorname{e}^x+1)- \ln(2)$. It can be seen as a smooth version of the more popular rectified linear unit.
In this description we omit the bias terms to reduce the notational clutter, but in the implementation a trainable bias vector is added after each matrix-vector product, i.e. there is an appropriately sized bias vector for each of the $W$'s.
As an edge update network we use a two layer neural network and the input is a concatenation of the sending and receiving vertex states and the current edge state.
\begin{align}
	e_{vw}^{t+1} &= E_t(h_v^t,h_w^t, e_{vw}^t) = g(W_|E2|^t g(W_|E1|^t(h_v^t; h_w^t; e_{vw}^t))),
	\label{eq:edge_update_func}
\end{align}
where $(\cdot; \cdot)$ denotes vector concatenation. This choice of edge update network means that the edge state for each of the two different directions between a pair of vertices become different after the first update.
This network is the only architectural difference from the SchNet model \cite{Schutt2018-ik}, i.e. if we set $e_{vw}^{t+1} = e_{vw}^{t}$ the model we describe here would be identical with the SchNet model.
The state transition function is also a two layer neural network. It is applied to the sum of incoming messages and the result is added to the current hidden state as in Residual Networks \cite{He2015-on}:
\begin{equation}
	S_t \left( h_v^t, m_{v}^{t+1} \right) = h_v^t + W_5^t g( W_4^t m_{v}^{t+1}),
\end{equation}
After a number of interaction steps $T$ we apply a readout function for which we use a two layer neural network that maps the vertex hidden representation to a scalar and finally we average over the contribution from each atom, i.e.
\begin{equation}
	R\left( \left\{ h_v^T \in G \right\} \right) = \frac{1}{N} \sum_{h_v^T \in G} W_{7} g(W_{6} h_v^T),
\end{equation}
In other words an atom and its chemical environment is mapped to an energy contribution.

\subsection{Initial Vertex and Edge Representation}\label{sec:init_representation}
The initial vertex hidden state $h_v^0$ depends on the atomic number of the corresponding atom.
The atomic number is used to look up a vector representation for that atom.
Using a hidden representation of size $256$ the initial hidden state is thus the result of a lookup function $\ell(x): \mathbb{N}\to \mathbb{R}^{256}$.
The weights in the vector representation are also trained during the optimization.

We use the model on three different levels of available information.
In the most ignorant scenario we have no labels on the edges of the graph and in this case the edge update function \eqref{eq:edge_update_func} just ignores the edge representation on the first layer, i.e. $e_{vw}^0$ is a ``vector'' of length 0 and $e_{vw}^t, t \in 1,\ldots,T$ are vectors of length $256$.
The next level of information is to include the point group information as described in section~\ref{sec:sym_classification}.
There are $9$ non-trivial point groups and we encode this information as an indicator vector of length 9, where $1$ means that the corresponding facet belongs to the given point group.
Finally we also run numerical experiments with the full spatial information for which the edges of the quotient graph are labeled with the interatomic distance. The distances are encoded by expanding them in a series of exponentiated quadratic functions as also done in \cite{Schutt2017-hk, Schutt2018-ik, Jorgensen2018-op}:
\begin{align}
	(e_{vw}^0)_{k} = \exp \left(-\frac{(d_{vw} - (-\mu_|min|+ k \Delta))^2}{2\Delta^2} \right),k=0\ldots k_|max|
	\label{eq:exponentiated_distance}
\end{align}
where $\mu_|min|$, $\Delta$, and $k_|max|$ are chosen such that the centers of the functions covers the range of the input features. This can be seen as a soft 1-hot-encoding of the distances, which makes it easier for a neural network to learn a function where the input distance is uncorrelated with the output.
In the experiments we use $\mu_|min|=0$, $\Delta=0.1$, and $k_|max|=150$.

\section{Datasets}\label{sec:datasets}

\begin{figure*}[htb!]
	\centering\includegraphics[width=1.0\textwidth]{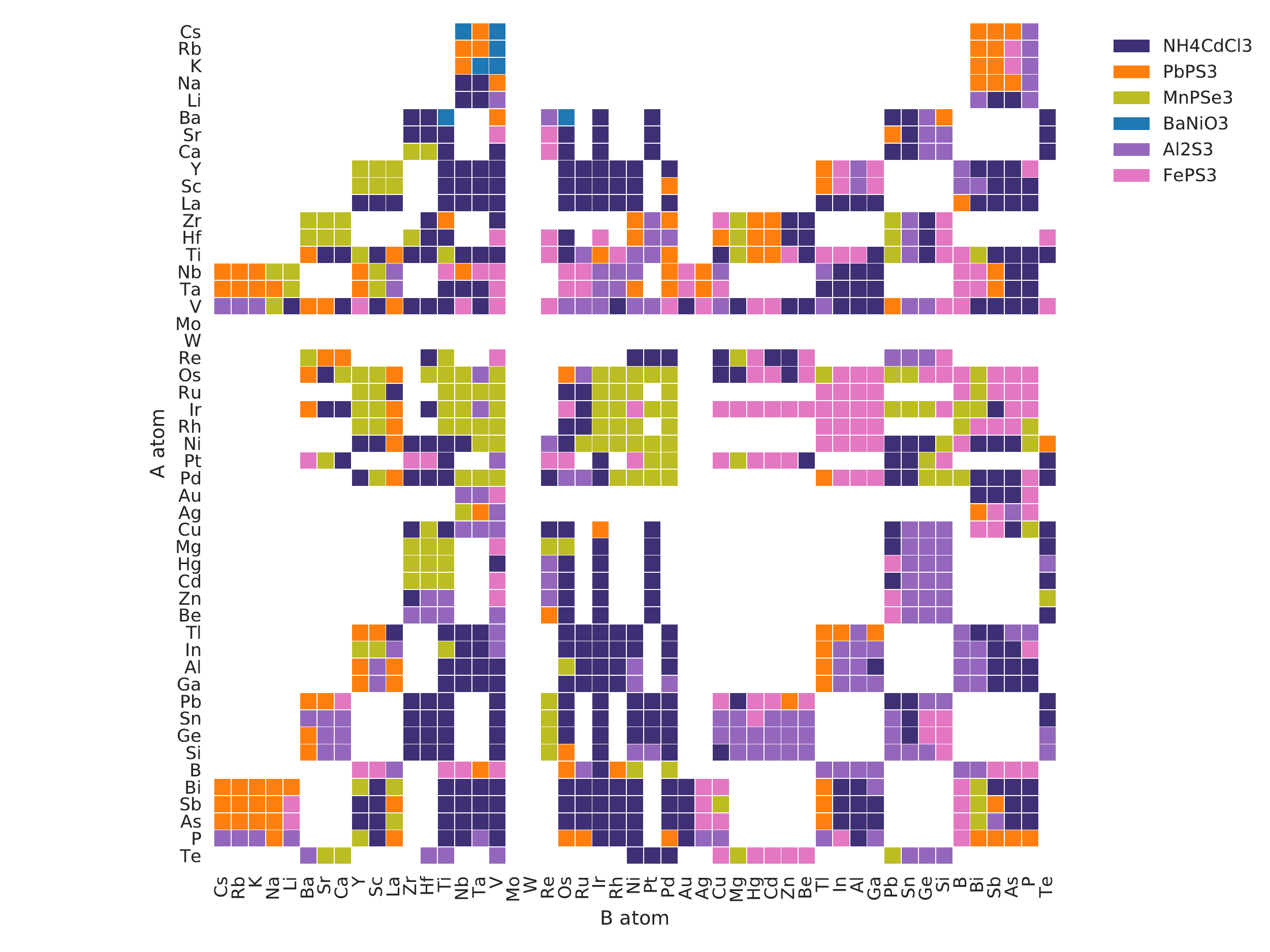}
	\caption{Map of the most stable prototype for each composition ABSe$_3$. The compositions that do not fulfill the valence rule have not been studied and thus, they are not colored.}
	\label{fig:abse3_heatmap}
\end{figure*}

For the numerical experiments we use two publicly available datasets and one dataset we generate.
\paragraph{The Materials Project \citep{Jain2013}} This dataset contains geometries and formation energies of \num{86680} inorganic compounds with input structures primarily taken from the  The Inorganic Crystal Structure Database (ICSD) \citep{Bergerhoff1983-me}. We use the latest version of the database (version 2018.11). The number of examples is reduced to \num{86579} after we exclude all materials with noble gases (He, Ne, Ar, Kr, Xe) because they occur so infrequently in the dataset that we consider them as outliers. This brings the number of different elements in the dataset down to 84.
\paragraph{Open Quantum Materials Database (OQMD) \citep{Saal2013-mi, Kirklin2015-xm}} Is also a database of inorganic structures and we use the currently latest version (OQMD v1.2) available on the project's website.
Again we consider materials with noble gases as outliers and we also exclude highly unstable materials with a heat of formation of more than \SI{5}{\electronvolt \per atom}.
Some entries in the database are marked as duplicates and we filter them in the following way: when a set of duplicates is encountered we use the first entry of the database, but if the standard deviation of the calculated heat of formation exceeds \SI{0.05}{\electronvolt \per atom} we discard the whole set of duplicates.
This leaves us with a total of \num{562134} entries.

For both datasets we split the entries into five parts of equal size to be used for 5-fold cross-validation, where the machine is trained on 4/5 of the data, and the remaining 1/5 is used for testing. For OQMD we also distribute the entries of OQMD that originates from ICSD equally between the five folds.

\paragraph{Ternary Selenides ABSe${}_3$}
For further testing, we have developed a third dataset of selenides. The intention behind this set is to test the ability of the model in a realistic computational screening setting. This dataset has only very limited overlap with OQMD, and predictions are made exclusively based on the symmetry labeled graphs of the new materials without any detailed information about the atomic coordinates.

The dataset contains the structures and formation energies of 5976 ternary selenides with stoichiometries ABSe${}_3$, where A and B are different transition metals in six different prototypes. 

The procedure for generating this dataset resembles the one presented in reference \cite{kuhar:2017}. We start by looking up the ABSe${}_3$ compounds reported in ICSD \cite{Bergerhoff1983-me}, and selecting the 6 prototypes that appear more than once:  hexagonal $P6_3/mmc$ structure of BaNiO${}_3$, orthorhombic $Pnma$ structure of NH${}_4$CdCl${}_3$/Sn${}_2$S${}_3$, monoclinic $C2/m$ FePS${}_3$, monoclinic $Pc$ structure of PbPS${}_3$, trigonal $R\bar{3}$ structure of MnPSe${}_3$ and hexagonal $P6_1$ structure of Al${}_2$S${}_3$. 

These structures are then used as templates, and we substitute the transition metal atoms A and B by 49 transition metals. Here, we avoid for simplicity Cr, Mn, Fe and Co, which usually lead to structures with large magnetic moments. We also limit ourselves to those combinations ABSe${}_3$ for which the valences of cations and anions add up to zero. This leads to a total of 512 ABSe${}_3$ compounds: 484 ternaries, which are then studied in 12 structures (6 for the ABSe${}_3$ and 6 for the BASe${}_3$) and 28 binaries, for which we study 6 different structures. A map to the compositions and structures studied can be found in figure \ref{fig:abse3_heatmap}.

The resulting 5976 structures have then been relaxed using Density Functional Theory (DFT) as implemented in the codes ASE \cite{ase-paper} and GPAW \cite{gpaw}. We perform two different kinds of electronic structure calculations: a coarse-grained calculation with the exchange-correlation functional PBEsol \cite{PBEsol} for the steps of the optimization and a fined-grained at the relaxed structure with the PBE exchange-correlation functional \cite{PBE}.  The cutoff energy for the plane wave basis set used to expand the wave functions is 800 eV in both cases. For the sampling of the Brillouin zone we use a Monkhorst–Pack mesh \cite{Monkhorst-Pack} with a density of 5.0/(Å${}^{-1}$) k-points in each direction for the relaxation steps and of 8.0/(Å${}^{-1}$) k-points for the refined calculation at the relaxed structure. All structures have been relaxed until the forces on the atoms are less than 0.05 eV/Å.

\section{Numerical Results and Discussion}\label{sec:results}
To assess the loss in accuracy going from full spatial information to unlabeled quotient graph we train/test the model in three different settings as mentioned in section~\ref{sec:init_representation}. In the most ignorant setting the quotient graph has only unlabeled edges. On the next level we label the edges with the symmetry of the corresponding Voronoi facet. With full spatial information the edges of the quotient graph are labeled with the distance between the atoms.
The model is trained with the Adam optimizer \cite{Kingma2014-tr} for up to \SI{10e6} steps using a batch size of 32. The initial learning rate is \SI{1e-4} and it is decreased exponentially so at step $s$ the learning rate is $10^{-4}\cdot 0.96^{\frac{s}{10^5}}$.
When training on OQMD and Materials Project we use \SI{5000} examples from the training data for early stopping. More specifically this validation set is evaluated every \SI{50000} steps and if the mean absolute error (MAE) has not improved for \SI{1e6} steps the training is terminated. When training on the ternary selenides ABSe${}_3$ dataset the 10\% of the training data is used as a validation set and the validation set is evaluated every training epoch.
In some of the results we use a model that has been pretrained on OQMD. In that case the model is trained on 4 out of 5 OQMD folds until the stopping criterion is met and the weights of the model are then used as initialization for training on the selenides dataset. The implementation of the model as well as the code used for generating the input graphs are available on Github \footnote{\url{https://github.com/peterbjorgensen/msgnet} \url{https://github.com/peterbjorgensen/vorosym}}.

\begin{table}[tbp]
	\centering
	\begin{tabular}{lrrrr}
		\toprule
	Dataset & Dist. & Sym & No sym & V-RF \\
		\midrule
OQMD all & 14 & 22 & 26 & 85\\
OQMD unary & 58 & 110 & 128 & 85\\
OQMD binary & 30 & 48 & 60 & 86\\
OQMD ternary & 14 & 20 & 23 & 80\\
ICSD all & 24 & 40 & 45 & 113\\
ICSD unary & 56 & 75 & 119 & 72\\
ICSD binary & 32 & 51 & 58 & 118\\
ICSD ternary & 22 & 35 & 39 & 109\\
Matproj all & 26 & 43 & 43 & 84\\
Matproj unary & 96 & 149 & 179 & 127\\
Matproj binary & 48 & 69 & 73 & 99\\
Matproj ternary & 27 & 43 & 43 & 87\\
		\bottomrule
	\end{tabular}
	\caption{MAE in meV/atom of test set energy predictions obtained through 5-fold cross-validation. The ICSD results are for the model trained on OQMD and tested only on the ICSD part of OQMD.}
	\label{tab:mae2}
\end{table}
\begin{table}[tbp]
	\centering
	\begin{tabular}{lrrrr}
		\toprule
		Dataset & Dist. & Sym & No sym & V-RF \\
		\midrule
OQMD all & 54 & 74 & 80 & 173\\
OQMD unary & 184 & 269 & 342 & 190\\
OQMD binary & 89 & 113 & 138 & 162\\
OQMD ternary & 52 & 70 & 71 & 131\\
ICSD all & 81 & 107 & 111 & 188\\
ICSD unary & 262 & 227 & 353 & 180\\
ICSD binary & 73 & 116 & 129 & 202\\
ICSD ternary & 88 & 112 & 102 & 182\\
Matproj all & 72 & 121 & 122 & 172\\
Matproj unary & 246 & 341 & 467 & 289\\
Matproj binary & 120 & 190 & 192 & 203\\
Matproj ternary & 65 & 119 & 111 & 181\\
		\bottomrule
	\end{tabular}
	\caption{RMSE in meV/atom of test set energy predictions obtained through 5-fold cross-validation. The ICSD results are for the model trained on OQMD and tested only on the ICSD part of OQMD.}
	\label{tab:rmse2}
\end{table}

\begin{figure*}[tbp]
	\centering\includegraphics[width=1.\textwidth]{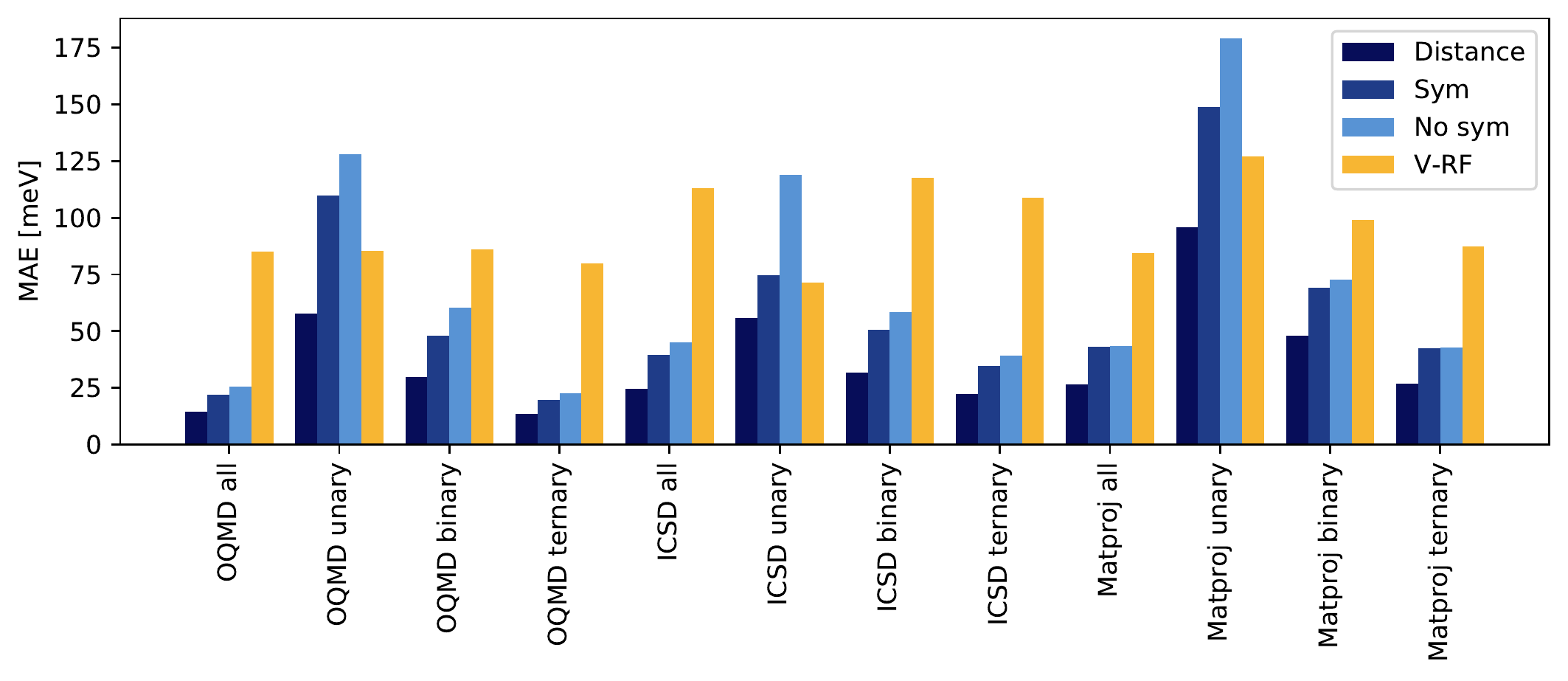}
	\caption{The figure shows the data in Table~\ref{tab:mae2}}
	\label{fig:barplot_mae}
\end{figure*}

\subsection{OQMD}
The mean absolute errors (MAE) and root mean squared errors (RMSE) of the test set predictions are shown in \tablename~\ref{tab:mae2} and the MAE is further visualized in Figure \ref{fig:barplot_mae}. As expected, the lowest prediction errors are obtained with the model where distance information is provided. If we focus on the OQMD the overall MAE is as low as 14 meV with distance information. This is lower than the SchNet-model \cite{Schutt2018-ik} by almost a factor of two because of the edge updates as discussed in Ref.~\cite{Jorgensen2018-op}. Two versions of the models without distance information are also shown. In one of them the symmetry information has not been used, but in the other one the symmetry classification of the Voronoi facets has been included as edge information. These two models do of course have larger errors than the one benefiting from the distance information, but still the error is surprisingly small. The MAE is only 22 meV for the model using symmetry information. For comparison the results for the model proposed by Ward et al. \cite{Ward2017-ff} is also shown in the figures (labeled V-RF for Voronoi - random forest). This model also builds on a Voronoi graph construction, but since the fractional areas of the Voronoi cells are provided, information about the distances are provided. Furthermore, many other attributes are added as information to the random forest algorithm applied. When this machine is applied to OQMD (using the same 5-fold splitting of the data as applied to the other algorithms) the resulting error is considerably larger, 85 meV, for all of OQMD.

To understand more about the behavior of the ML algorithms investigated here, we have considered the test errors on different subsets of OQMD and also on the Material Project database \cite{Jain2013}. Let us first note that the OQMD contains two different types of structure sources. One type, which gives rise to the largest number of materials, consists of a number of fixed crystal structures or prototypes decorated by the different chemical elements. There are 16 elemental prototypes, 12 binary ones, and 3 ternary ones. For two of the ternary ones, one of the elements is predefined to be oxygen. This generates a very large number of materials of varying composition and stability, but in a fairly small number of different crystal structures. The other type of structures comes from materials from the experimental ICSD database. This group is characterized by a much greater variation in the crystal structures, but is naturally limited to mostly stable materials, since they have been experimentally synthesized.

We first consider the test error on the subsets of OQMD consisting of the unary, binary, and ternary systems, and we shall focus on the model where the symmetry information is included, but the distances are not. As can be seen from Table~\ref{tab:mae2}, the test error is considerably larger on the unary systems (110 meV) than on the database as a whole. This also holds for the binary ones but to a smaller degree (48 meV). It is not clear to us at the moment exactly why this is so, but we shall discuss some possible explanations. The unary and binary systems only constitute a fairly small part of the total database, and the weight of these systems during the training is therefore also limited. Another factor may be that a large fraction of the unary and binary systems belong to the group of materials where the crystal structures are systematically generated as mentioned above. This means that many rather "artificial" and unstable materials are generated, where the atoms are situated in environments, which do not occur in reality, and the resulting energies may be far above more stable structures. This could potentially be difficult for the machine to learn.

\subsection{ICSD/OQMD}
Table~\ref{tab:mae2} also shows the results for the ICSD subset of the OQMD database. The results shown are for the model trained on all of OQMD but tested only on the ICSD subset. The overall MAE is seen to be roughly a factor of two larger than for all of OQMD. This is probably due to the fact that the ICSD is a subset with a large variation of structures and this makes prediction more difficult on average. We see the same trend as for all of OQMD, that the error decreases going from unary to binary to ternary systems. For the unary systems the test error is in fact lower for the ICSD subset than for all of OQMD, which may be due to the fact that physically artificial high-energy systems appear in OQMD but not in ICSD. For the binary systems there is a balance: the ICSD does not contain so many high-energy systems, which could make predictions better, but on the other hand the larger variation of crystal structures is more difficult to predict.

\subsection{Materials Project Database}
The models have also been trained and tested on the Materials Project dataset \citep{Jain2013}. The overall error is fairly similar to the one obtained for the ICSD subset of OQMD as might be expected since the Materials Project is also based on mostly materials from the ICSD. The errors for the unary and binary subsets are somewhat larger for the Materials Project database. This might be due to the fact that the machine trained on OQMD benefits from the larger number of systematically generated unary and binary systems in that database.

\subsection{RMSE vs. MAE}
The root-mean-square-errors are shown in Table~\ref{tab:rmse2}. In all cases the values are considerable higher than the MAE. This is an indication that the distribution of the errors have heavier tails than a Gaussian, and as we shall see in the following examples that a significant number of outliers exist. The outliers might be due to limitations of the model but could also appear because of problematic entries in the database as also discussed by Ward et al.~\cite{Ward2017-ff}.

\subsection{Solid angle cutoff of Voronoi facets}

\begin{figure}[tbp]
	\centering\includegraphics[width=0.8\columnwidth]{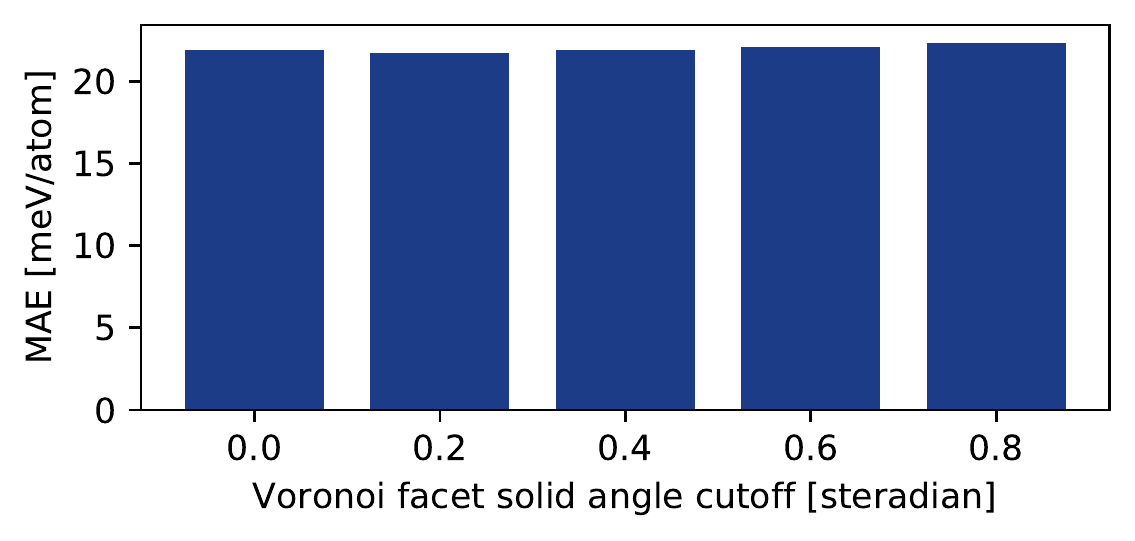}
	\caption{Prediction error on OQMD test set vs Voronoi facet solid angle cutoff $\Omega_{cut}$ for the model using symmetry labels. The error decreases slightly when removing small facets and increases only slowly when $\Omega_{cut}$ is further increased.}
	\label{fig:cutoff_sweep}
\end{figure}

The above results are all calculated using a cutoff of the Voronoi facet solid angle of $\Omega_{cut} = 0.2$. However, the results are almost independent of the value as shown in Figure~\ref{fig:cutoff_sweep}, where the MAE on all of OQMD is shown for the model where symmetry but no distance information is included. We see that the error decreases slightly when small facets are removed with $\Omega_{cut} = 0.2$, and increases only slowly when $\Omega_{cut}$ is further increased. We take this as an indication that the connectivity of the material is well described even when the graph is reduced to essentially include only nearest neighbor bonds.

\subsection{ABO$_3$ materials in OQMD}

\begin{figure}[tbp]
	\centering\includegraphics[width=1.0\columnwidth]{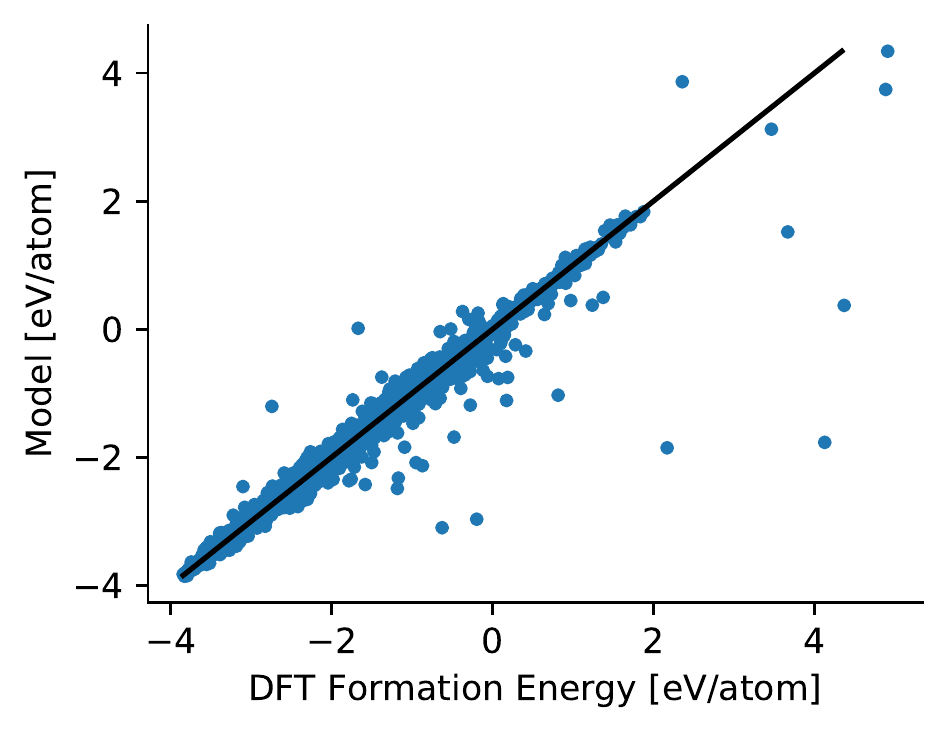}
	\caption{Test set predictions on 12395 ABO$_3$ structures of OQMD (MAE=36, RMSE=112 \si{m\electronvolt \per atom}) using 5-fold cross-validation, i.e. the plot is a collection of predictions from 5 different models, each trained on 4/5 of the data and tested on the remaining 1/5.}
	\label{fig:abo3_predictions}
\end{figure}

We now consider the subset of all oxides in the OQMD with the composition ABO$_3$. We shall investigate to which extent the model is able to predict the right ground state structure for a given composition. We first show the overall prediction for the 12935 materials of this type in OQMD in Figure~\ref{fig:abo3_predictions}. We again use the model with symmetry-labeled graphs without distance information. The MAE is 36 meV, which is about the same value as the one for the subset of ternaries in ICSD (35 meV). The RMSE is again significantly higher (112 meV) because of severe outliers as can be seen in the plot.

\begin{figure}[htbp]
	\centering\includegraphics[width=1.0\columnwidth]{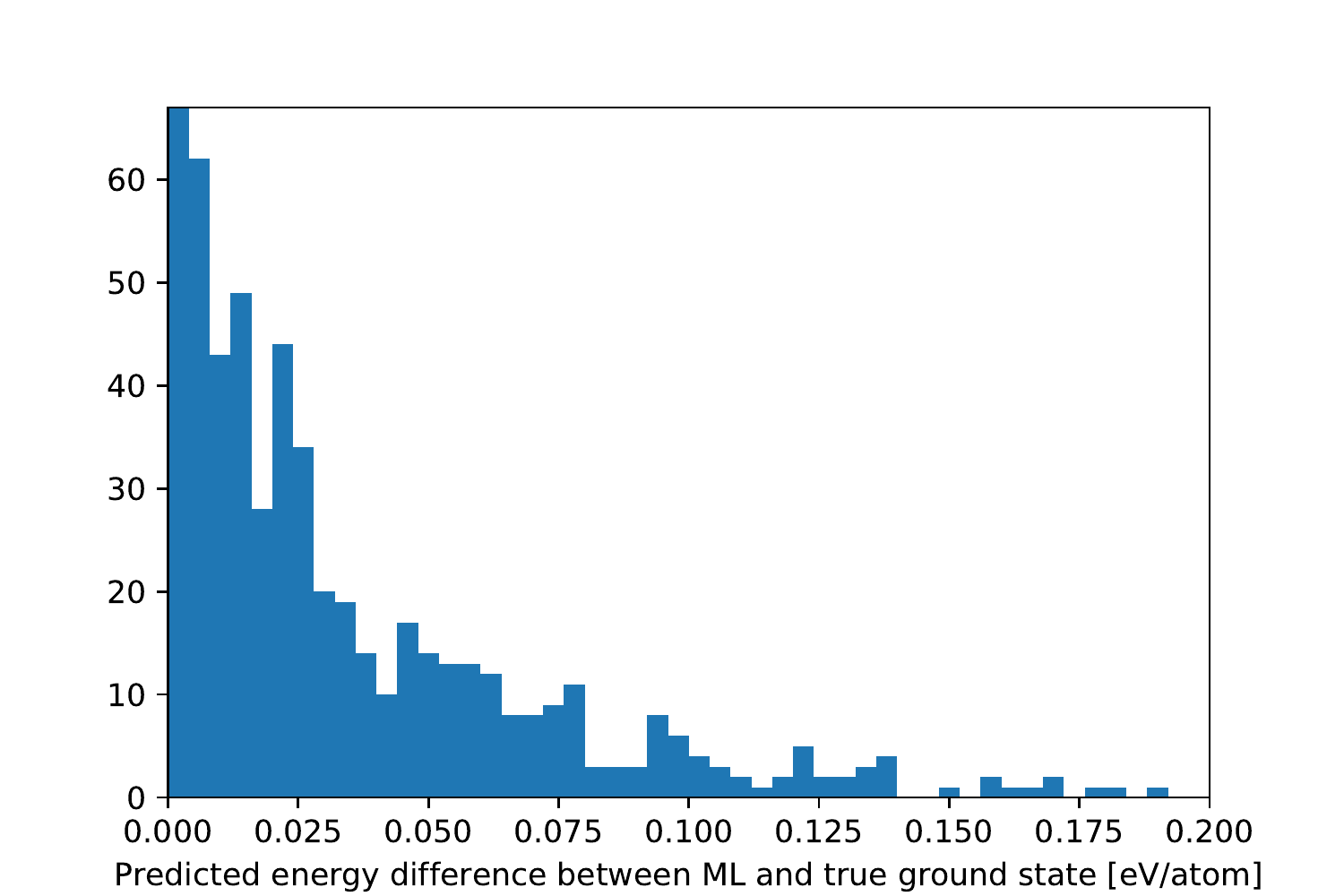}
	\caption{Predicted energy difference between the DFT ground state and the ML ground state: $\Delta E = E^{ML}(G_\textrm{DFT}) - E^{ML}(G_\textrm{ML})$ for the ABO$_3$ materials in OQMD. The total number of compositions is 2646. The peak at zero is much higher than shown in the graph. It corresponds to the 2097 compositions, where the right ground state is predicted. For the remaining 549 compositions the mean absolute difference is \SI{44}{m\electronvolt \per atom}.}
	\label{fig:abo3_structural}
\end{figure}

\begin{figure*}[tb]
	\subfloat[Predicted energy for selenides in OQMD (MAE=24 RMSE=38 \si{m\electronvolt \per atom})]{\centering\includegraphics[width=.29\textwidth]{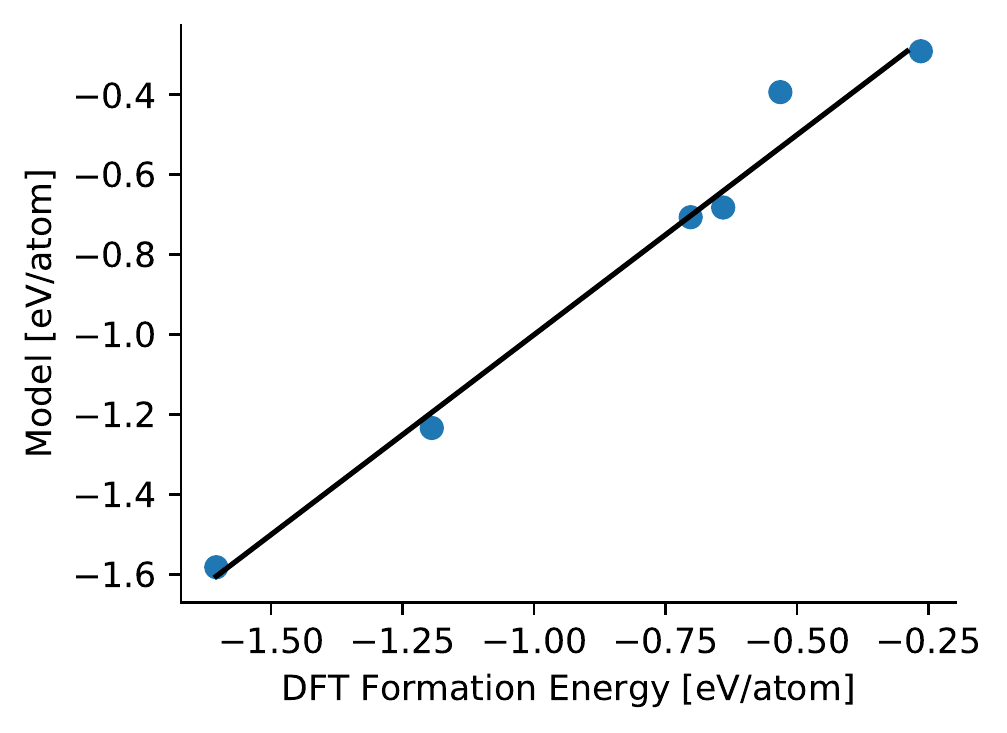}}
	\subfloat[Predicted energy for selenides (initial structures) using model trained on OQMD (MAE=176 RMSE=236 \si{m\electronvolt \per atom})]{\centering\includegraphics[width=.29\textwidth]{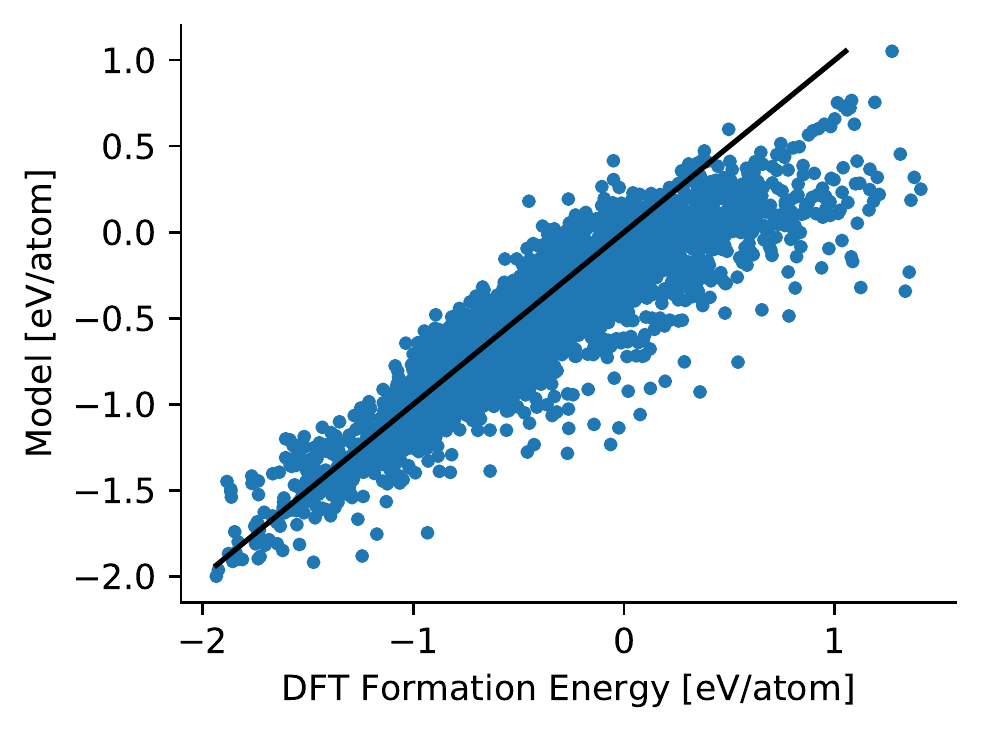}}
	\subfloat[Predicted energy for selenides (initial structures) using model pretrained on OQMD and then on 100 selenides (MAE=95 RMSE=129 \si{m\electronvolt \per atom})]{\centering\includegraphics[width=.29\textwidth]{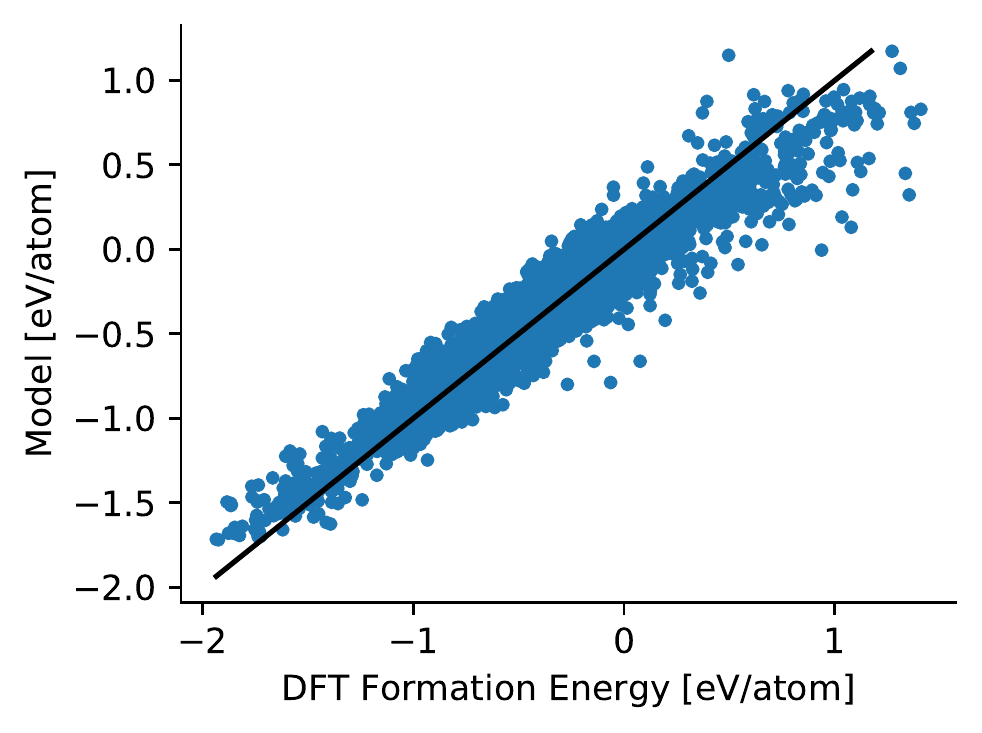}}
	\caption{Predictions on ABSe$_3$ test set with model pretrained on OQMD.}
	\label{fig:abse3_prediction}
\end{figure*}

We now ask the following question: given a composition (A,B) the model predicts a ground state structure $G_\textrm{ML}$. If we are going to investigate this structure and other low energy structures with DFT, how high up in energy (as predicted by the model) do we have to go before we find the DFT ground state structure $G_\textrm{DFT}$? We only include entries for which there is more than one structure (12329/12395) and the average number of structures per composition is 4.7 The energy difference $\Delta E = E^{ML}(G_\textrm{DFT}) - E^{ML}(G_\textrm{ML})$ of course varies from system to system, and the distribution is shown in Figure~\ref{fig:abo3_structural}. The mean absolute difference (MAD) of this distribution is very small, only 9 meV, and the maximum error is a clear outlier at 0.92 eV. The reason for the small MAD is that for 2097 out of the 2646 compositions the correct ground state is predicted, however, in many cases because only two structures exist in the database for a given composition. For comparison the expected number of correctly predicted ground state structures with random guessing is 843. If we only look at the 549 compositions for which the ML model predicts the wrong ground state the MAD is \SI{44}{m\electronvolt \per atom}. For comparison the energy prediction for the ground state structures has an MAE of \SI{29}{m\electronvolt \per atom}.
The low MAD value of 44 meV is promising for applications to computational screening. It sets an energy scale for how many structures have to be investigated by DFT to identify the DFT ground state after the model predictions have been generated.

\subsection{ABSe$_3$ selenides}

The last dataset we shall consider consists of selenides with the ABSe$_3$ composition as discussed in the section about the datasets. This dataset is considerably more challenging for two reasons. Firstly, there is very little overlap between this dataset and the training dataset OQMD. Only 6 materials are shared between the two datasets, and the test predictions for these are shown in Figure~\ref{fig:abse3_prediction}a. The MAE is 24 meV, and the RMSE is also low, only 38 meV. The second challenge is, that we shall now use the model to make predictions based on the initial graph before relaxations. The 6 different prototypes in the dataset each have a graph in the original material giving rise to the naming of the prototype. For example, one of the types is hexagonal $P6_3/mmc$ structure of BaNiO${}_3$, so for predictions in this structure we shall use the graph of BaNiO$_3$. Some of the prototype structures have a fair number of atoms in the unit cell (up to 20) and a low symmetry (monoclinic), which means that there are many free atomic coordinates that are optimized during relaxation. This leads to frequent modifications of the graph during relaxation.

Figure~\ref{fig:abse3_prediction}b shows the model predictions based on the initial prototype graphs versus the DFT energies of the resulting optimized structures. The MAE is 176 meV, which is considerably higher than the value for the oxides. Particularly large deviations are seen for large and positive heats of formation. In a computational screening setting this might not be an issue because the high-energy materials are going to be excluded anyway. The RMSE is only a factor 236/176 = 1.34 larger than the MAE, which is due to the small number of outliers compared to for example the oxides (Figure~\ref{fig:abo3_predictions}).  

The prediction quality can be significantly improved by additional training on the selenide dataset. Even a limited number of data points have a considerable effect. This is to be expected since the overlap between the selenide dataset and the OQMD is only 6 materials as mentioned above. Figure \ref{fig:abse3_prediction}c shows the model-DFT comparison if the model is first trained on the OQMD dataset and then subsequently trained on 100 materials out of the 5976 selenides in the database. The MAE is reduced from 176 meV to 95 meV bringing the error down to a value comparable to the error between DFT and experiment \cite{Kirklin2015-xm}.

\begin{figure}[htbp]
	\centering\includegraphics[width=0.9\columnwidth]{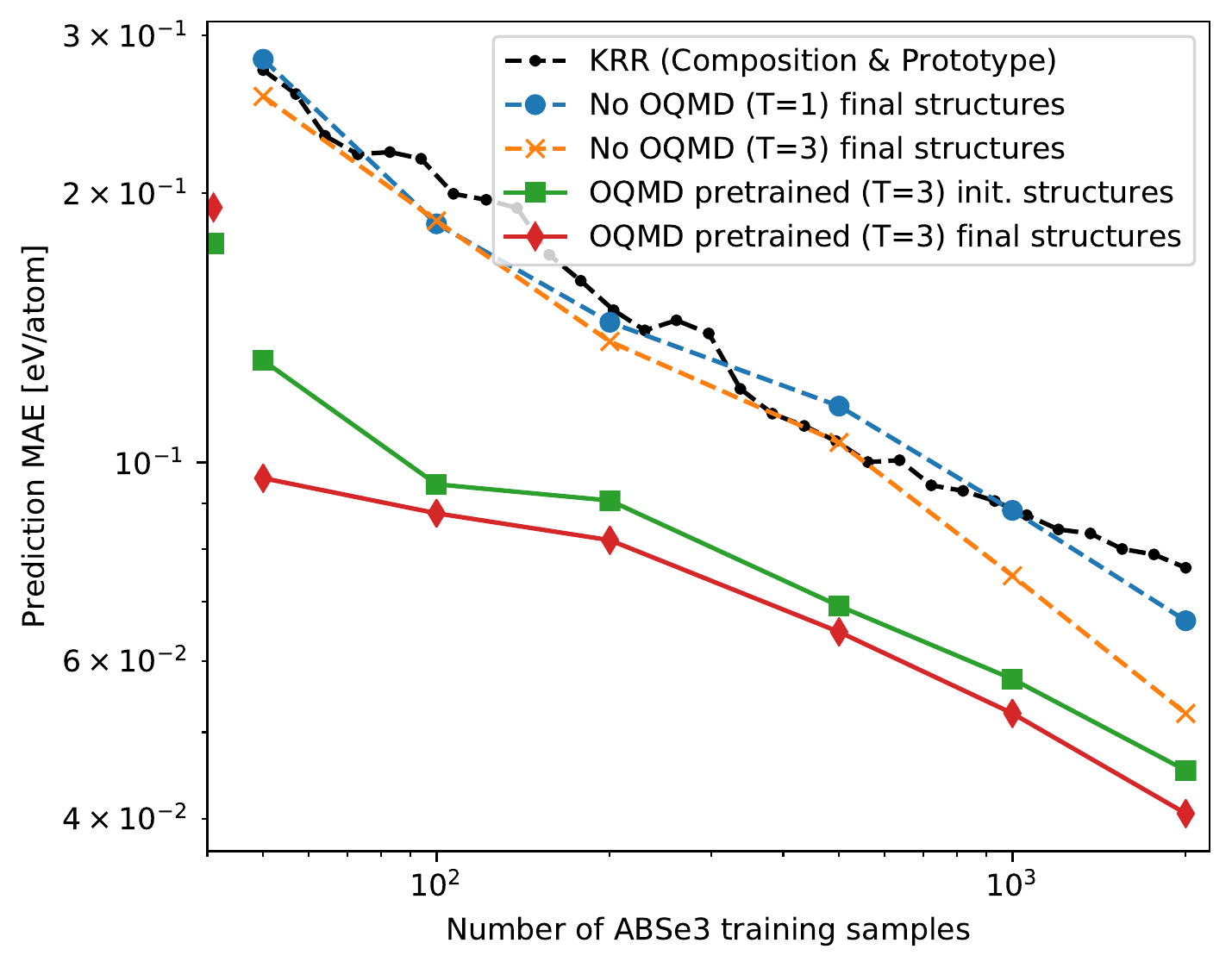}
	\caption{Predictions on ABSe$_3$ structures with increasing number of ABSe$_3$ training samples. The solid lines correspond to models that have been pretrained on OQMD and then on the ABSe$_3$ dataset. The unconnected points correspond to the model only trained on OQMD final structures, i.e. the pretrained model. The parameter $T$ denotes the number of interactions steps and initial/final structures refers to whether the model input is the graph derived from the prototype structure or the DFT-relaxed structure. KRR denotes a kernel ridge regression baseline model using only composition and prototype as input.}
	\label{fig:abse3_learning_curves}
\end{figure}

The effect of additional training on the selenide dataset is shown as a function of training set size on a logarithmic scale in Figure~\ref{fig:abse3_learning_curves}. The points on the $y$-axis correspond to the situation without any additional training. As can be seen, a small amount of additional training leads to significant reduction of the prediction error. The solid curve with square markers corresponds to the situation discussed above where the model is first trained on OQMD, and then further trained on the initial graphs (but relaxed energies) for part of the selenides. For comparison, the solid curve with diamond markers shows the prediction error, when the training and prediction is based on the final graph. Using the initial graphs instead of the final graphs gives rise to only a slightly higher MAE. This is encouraging for the potential use of the approach in computational screening studies, where predictions have to be based on the initial prototype structures to avoid the computationally costly DFT calculations. 

As a baseline we also show the results of the model if it is trained exclusively on the selenide dataset (dashed curve with cross markers). As expected the MAE is much larger than for the pretrained model for small amounts of data. For larger training sets the MAE drops gradually and with a dataset size of about 500 materials the prediction error is comparable to the one for the OQMD-pretrained model, which is trained on an additional 50 selenides. We ascribe the rather successful performance of the model without pretraining at large training set sizes to the systematic character of the dataset: only 6 different crystal structures are represented and they are systematically decorated with a particular subset of atoms. The last model (dashed curve with circle markers) is again only trained on the selenide dataset, but now only one interaction step ($T=1$) is performed in the message passing neural network in contrast to the three iterations used otherwise. The performance is seen to be rather similar to the model with $T=3$ up to a training dataset size of 300. With only one iteration in the network information about the identity of neighboring atoms is exchanged, and this is apparently sufficient to roughly characterize the 6 crystal structures. At larger training set sizes, where the prediction error is smaller, the network with three iterations outperforms the one with only one iteration.

We also include an even simpler baseline model that uses only the composition and the prototype as input and is only trained on the ABSe$_3$ dataset.
For this baseline the input vector representation consists of a 1-hot-encoding for the atom type of the A-atom, a 1-hot-encoding for the B-atom and a 1-hot-encoding for the prototype.
We use kernel ridge regression (KRR) as implemented by scikit-learn using the RBF-kernel and using 10-fold cross-validation to choose the hyper-parameters $\alpha$ ($\ell2$-penalty weight) and $\gamma$ (kernel length-scale) on the grid $\alpha \in [1, 0.1, 0.01, 0.001]$ and $\gamma \in [0.01, 0.1, 1.0, 10.0, 100.0]$. The prediction error is similar to the other baseline model that uses only one interaction step.

\begin{figure}[htbp]
	\centering\includegraphics[width=1.0\columnwidth]{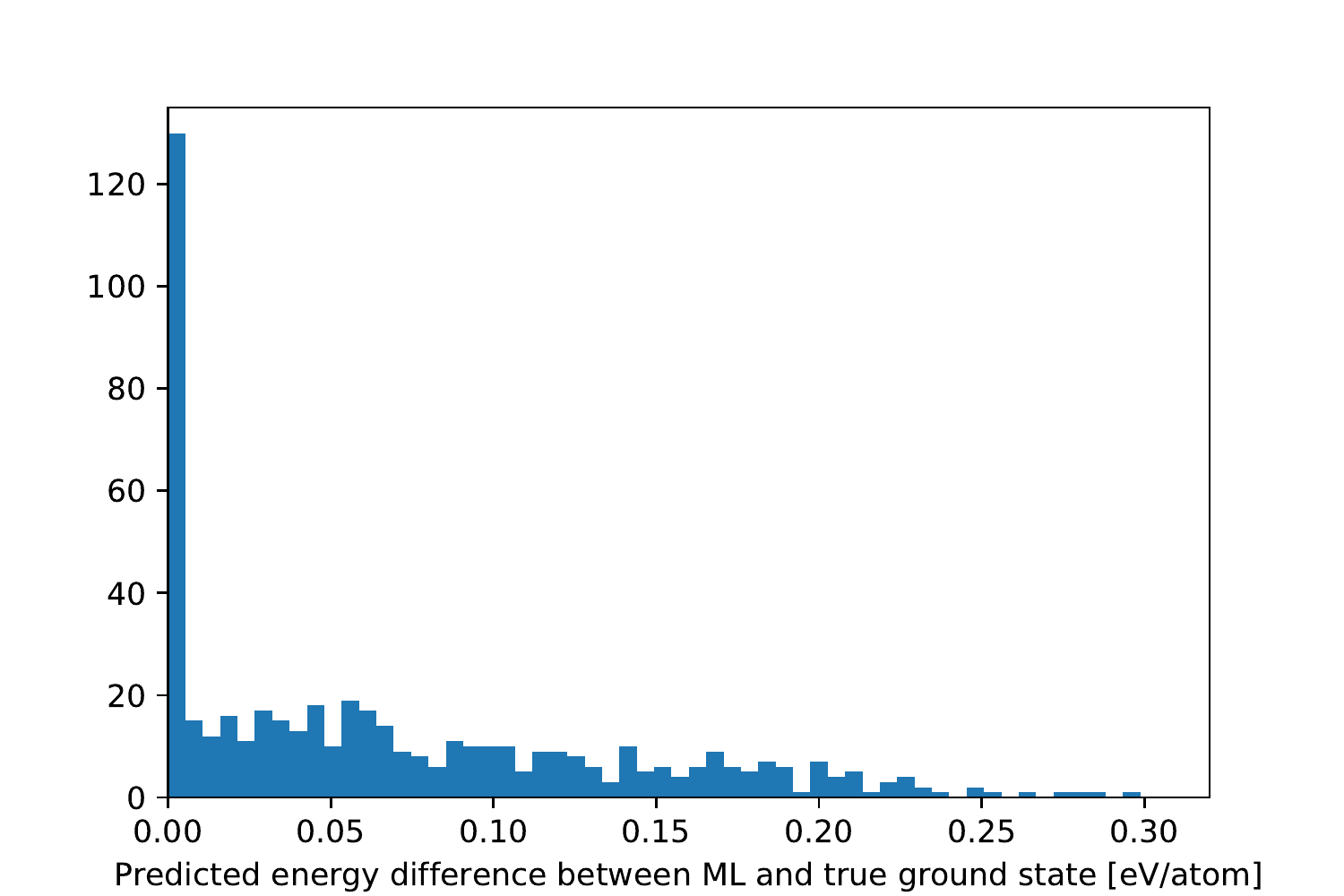}
\caption{Energy difference between the ML-predicted ground state and the true ground state, $\Delta E = E^{ML}(G_\textrm{DFT}) - E^{ML}(G_\textrm{ML})$, for the selenide dataset. The mean absolute difference is \SI{70}{m\electronvolt\per atom}.}
	\label{fig:abse3_structural}
\end{figure}

Figure~\ref{fig:abse3_structural} shows the distribution of the predicted energy difference between the DFT ground state structure and the ML predicted ground state structure, $\Delta E = E^{ML}(G_\textrm{DFT}) - E^{ML}(G_\textrm{ML})$ for the selenide dataset. Only in 104 out of the 512 compositions, the model predicts the DFT ground state. This in not particularly impressive, since random prediction of a structure would give roughly $512/12 \approx 43$ correct predictions. However, the dataset have many low-lying energy structures, where even full DFT calculations cannot be expected to necessarily predict the correct ground state structure. This was investigated in more detail in a similarly generated dataset of ABS$_3$-sulfides used for computational screening of water-splitting materials \cite{kuhar:2017}. The mean absolute difference is only \SI{70}{m\electronvolt\per atom} with a maximum error of \SI{0.3}{\electronvolt \per atom}. The low mean value is clearly promising for future applications to computational materials screening.

\section{Conclusions}\label{sec:conclusions}
In summary, we have proposed a ML model for the prediction of the formation energy of crystalline materials based on Voronoi quotient graphs and a local symmetry description. It uses a message passing neural network with edge updates. The model is independent of the detailed atomic positions and can therefore be used to predict the formation energy of new materials, where the detailed structure is unknown. 

The model test MAE is very small (22 meV) on the OQMD dataset, and a factor of two larger (40 meV) on the ICSD subset of OQMD.
To test the model in a realistic materials screening setting, we created
a dataset of 6000 selenides with very small overlap with the OQMD.  The model pretrained on OQMD and applied to the selenides shows an MAE of 176 meV. This value can be lowered to 95 meV with an additional training on 100 selenides. Further training can lower the MAE to below 50 meV.

Based on the results we conclude, that is possible to develop ML models with position-independent descriptors, which are useful for realistic materials screening studies. However, extrapolation from OQMD to other datasets is challenging. One reason for this may be, as pointed out before, that the OQMD is composed of materials of two types: Some are generated systematically in rather few predefined crystal structures while others come from ICSD. (There is of course a significant overlap between the two types). The first type is characterized by a large variation in stability, but low variation in crystal structures, while the second type is the opposite: the experimentally observed materials in ICSD exhibit a large variation in crystal structures, but they are all (except for some high-pressure entries) stable low-energy materials. This bias might limit the extrapolation to datasets which contain structures weakly represented in OQMD and with element combinations, which are far from stable. One way forward could be to create datasets with less bias so that unstable materials are represented in a greater variety of structures.

We see a number of potential improvements of the proposed model. More symmetry information could be included using for example Wyckoff positions \cite{Jain2018-df} or additional graph edges describing symmetry relations. Furthermore, it is possible to label the quotient graphs with crystal translation information so that the infinite graph can be reconstructed \cite{Klee:2004cw}. This would make the crystal description more unique. 

Perhaps the model could also learn the atomic positions from the graph representation.
The latest developments in generative models have succeeded in generating small molecules in 3D space \cite{Gebauer2018-bc}. By combining this kind of model with the restrictions imposed by the connectivity and symmetries described by the quotient graph (see for example \cite{Thimm2009-fs, Eon2011-fx}) it might be possible to directly predict the atomic positions without running DFT relaxations.

Another useful extension would be to model uncertainties in the prediction. Even though the datasets used here have a relatively high number of entries they only contain a tiny fraction of the chemical space.
If the model could learn what it does not know it would be very useful in an active learning setting where DFT calculations could be launched by the model to explore areas of the chemical space with high uncertainty.
A promising direction for uncertainty modeling is to use ensembles of neural networks where different techniques can be considered to ensure diversity between ensemble members \cite{Peterson2017-mv, Lakshminarayanan2017-eh,Pearce2018-rm,Osband2018-pl}.

\section{Acknowledgments} We would like to thank Peter Mahler Larsen for helpful discussions. We acknowledge support from the VILLUM Center for Science of Sustainable Fuels and Chemicals which is funded by the VILLUM Fonden research grant (9455) and thanks to Nvidia for the donation of one Titan X GPU.

\bibliography{main}
\end{document}